# Exceeding the limits of algorithmic self-calibration in super-resolution imaging

**Eric Li,**[1,*] **Stuart Sherwin,**[1] **Gautam Gunjala,**[1] **and Laura Waller**[1]

[1]*Department of Electrical Engineering and Computer Sciences, University of California, Berkeley, CA 94720, USA*
*\*esli@berkeley.edu*

**Abstract:** Fourier ptychographic microscopy is a computational imaging technique that provides quantitative phase information and high resolution over a large field-of-view. Although the technique presents numerous advantages over conventional microscopy, model mismatch due to unknown optical aberrations can significantly limit reconstruction quality. Many attempts to address this issue rely on embedding pupil recovery into the reconstruction algorithm. In this paper we demonstrate the limitations of a purely algorithmic approach and evaluate the merits of implementing a simple, dedicated calibration procedure. In simulations, we find that for a target sample reconstruction error, we can image without any aberration corrections up to a maximum aberration magnitude of $\lambda/40$. When we use algorithmic self-calibration, we can increase the aberration magnitude up to $\lambda/10$, and with our *in situ* speckle calibration technique, this working range is extended further to a maximum aberration magnitude of $\lambda/3$. Hence, one can trade-off complexity for accuracy by using a separate calibration process, which is particularly useful for larger aberrations.

## 1. Introduction

Fourier ptychographic microscopy (FPM) is a well-known method that captures image data from multiple illumination angles to reconstruct a high-resolution image that achieves the resolution of a high-NA objective with the FOV of a low-NA objective [1–4]. In addition to providing resolution enhancement, FPM recovers quantitative phase information from intensity-only measurements of the sample. However experimental errors such as illumination misalignment and imaging aberrations have been shown to degrade the performance of FPM, both qualitatively and quantitatively [1, 5, 6].

One strategy for dealing with unknown aberrations in the imaging system is known as embedded pupil recovery (EPRY), which modifies the reconstruction algorithm to simultaneously recover both the unknown object and the pupil (aberration) function using the same dataset [5, 7, 8]. This method is purely algorithmic and extends the practical settings under which quantitatively accurate FPM is possible. However, the performance of this method is limited by the interdependence of object and pupil reconstructions. Images of distinct samples acquired using the same system do not necessarily reveal the same pupil function using EPRY.

An alternative approach to correcting unknown aberrations is to measure the aberrations in advance using a separate set of calibration images, and subsequently using them to correct for aberrations in the FPM reconstruction. Specifically, we consider using a speckle imaging approach, which only requires placing a weakly scattering diffuser at the sample plane of the imaging system [9]. Aberrations are recovered from the power spectra of acquired speckle images. By basing the aberration estimation on a separate set of speckle images, we obtain an unbiased estimate of the pupil function, which in turn provides an unbiased reconstruction of the sample. This *in situ* metrology technique is readily applicable to FPM, for which an imaging system must already have steerable illumination, and it only requires the additional acquisition of roughly 10 speckle images.

There is particular interest in implementing FPM as a practical solution for accurately measuring the phase of EUV photomasks at the operating wavelength of 13.5 nm, which is of critical importance for EUV lithography [10–13]. Characterizing the maximum error tolerance is crucial for porting FPM to synchrotron imaging systems [14], which typically operate at wavelengths shorter than visible light, where the impact of wavefront errors is magnified.

We model an extreme ultraviolet (EUV) imaging system based on the SHARP (SHARP High-NA Actinic Reticle review Project) microscope at the Advanced Light Source at Lawrence Berkeley National Laboratory. SHARP is an EUV microscope providing diffraction limited imaging of EUV photomasks to emulate the conditions of an EUV lithography scanner used in leading-edge semiconductor manufacturing. SHARP uses steerable plane wave illumination at a wavelength of 13.5nm coupled with a single off-axis zone plate to produce an image of the mask's reflection at a resolution of $\lambda/2NA \approx 82$nm [15]. However images from SHARP are corrupted by aberrations that increase with distance from the center of the FOV [16, 17], degrading the FPM reconstruction accuracy unless properly accounted for in the transfer function. Accurately characterizing the field-vary aberrations could greatly increase the usable FOV on SHARP, thus reducing the need for spatial scanning, allowing for increased throughput and decreased operational cost. Our diffuser-based method enables accurate object reconstruction at higher aberration magnitudes than self-calibrated super-resolution imaging.

In this work, we simulate the performance of these two methods for FPM under the presence of unknown aberrations of varying magnitude. For each method we identify the maximum magnitude of aberrations under which the complex field of the object can be reconstructed with under 10% error. In the absence of any aberration correction strategies, we find this level of accuracy can only be attained when aberrations are smaller than $\lambda/40$. In contrast we find that the purely algorithmic approach of EPRY extends the maximum tolerable aberration magnitude to $\lambda/10$, an increase of 4× compared with no calibration. Finally, the speckle imaging approach further extends the maximum aberration magnitude to $\lambda/3$, an additional increase of 3.3× compared with self-calibration, and 13× compared with no calibration. For SHARP, the aberration magnitude over a 5um path ranges from under $\lambda/21$ near the center of the image, to over $\lambda/2$ near the edge [18]. Therefore on this system, calibration-free FPM (requiring $< \lambda/40$) would be unable to attain quantitative accuracy over any 5um patch in the image; EPRY (requiring $< \lambda/10$) could enable quantitative accuracy close to the center of the FOV; and the proposed speckle imaging calibration (requiring $< \lambda/3$) could extend the usable FOV to encompass nearly the entire imaging sensor.

We structure our paper as follows: In Section 2, we define our optical system model and summarize both aberration recovery algorithms. In Section 3, we outline our simulation procedure and present comparative results for both reconstruction techniques considered. In section 4, we discuss the how these results clearly demonstrate the value of independent aberration metrology for moderately and highly aberrated systems.

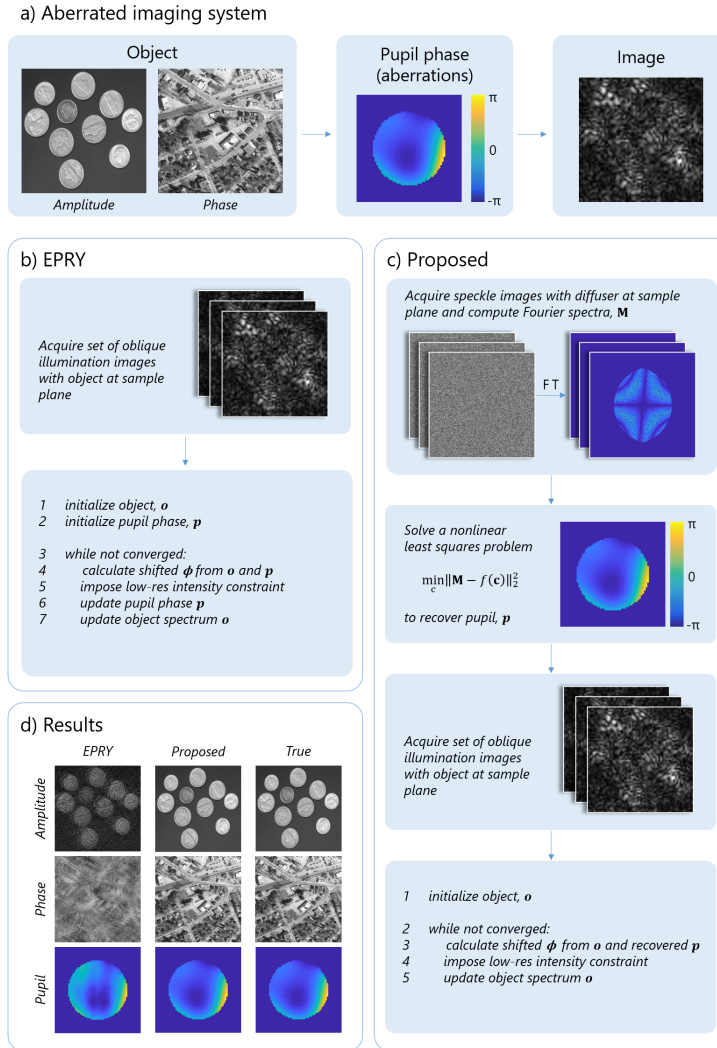

Fig. 1. We examine the difference in Fourier ptychographic microscopy (FPM) reconstruction quality in aberrated systems using purely computational calibration (Embedded Pupil Recovery, EPRY) and a speckle imaging calibration step. (a) Images acquired using an imaging system with $\lambda/5$ waves-rms of wavefront error (aberrations) are simulated for FPM reconstruction. (b) In EPRY, the object and pupil functions are jointly optimized from a single set of aberrated measurements. For aberrations of this magnitude, EPRY suffers from significant artifacts such as crosstalk between phase and amplitude, as well as low-frequency phase errors. (c) For speckle imaging calibration, we collect an additional set of images of a weakly scattering diffuser under varying angles of illumination. These images are used to recover the pupil function only. The pupil estimate is then inserted into the FPM algorithm, which solves for the sample only. (d) Our proposed diffuser pre-calibration achieves a much more faithful reconstruction than EPRY for both pupil and object, with minimal reconstruction artifacts.

## 2. Optical system model

We model an imaging system with the required illumination for FPM - spatially coherent, steerable plane-wave illumination [1]. We can compute the image from source point $\boldsymbol{u} = \frac{1}{\text{NA}}\left(\sin\theta_x, \sin\theta_y\right)$ by applying the optical transfer function to the incident electric field [19], where $\mathcal{F}$ is the Fourier operator and $\mathcal{F}^{-1}$ is its inverse:

$$I_{\text{out}}(\boldsymbol{x}) = \left|\mathcal{F}^{-1}\left[\hat{P}(\boldsymbol{u}) \cdot \mathcal{F}[E_{\text{illumination}}(\boldsymbol{x})O(\boldsymbol{x})]\right]\right|^2. \tag{1}$$

The term $E_{\text{illumination}} = \exp\left(-i2\pi\frac{\text{NA}}{\lambda}\boldsymbol{u}\cdot\boldsymbol{x}\right)$ describes the incident plane-wave that enters the optical system with spatial frequency $\frac{\text{NA}}{\lambda}\boldsymbol{u}$. The term $O(\boldsymbol{x})$ represents the unknown object, and $\hat{P}(\boldsymbol{u})$ represents the optical transfer function of the imaging system. We assume the following transfer function $\hat{P}(\boldsymbol{u})$ represents an ideal circular aperture with a phase aberration:

$$\hat{P}(\boldsymbol{u}) = \exp\{i \cdot W(\boldsymbol{u})\} \cdot \text{circ}(\boldsymbol{u}). \tag{2}$$

The magnitude of the transfer function is modeled as a binary indicator function, $\text{circ}(\boldsymbol{u})$, owing to the approximately uniform attenuation across all frequencies within the aperture of standard zone-plate lenses in SHARP [20]. The system aberrations manifest in the phase of the transfer function $W(\boldsymbol{u})$, which we refer to as the wavefront error function (WEF).

### 2.1. Fourier ptychographic microscopy

Fourier ptychographic microscopy is a synthetic aperture microscopy technique that enables super-resolution imaging with quantitative phase retrieval. It is performed by acquiring multiple images of the sample at a series of illumination angles and solving an optimization problem to recover quantitative amplitude and phase. We perform FPM reconstructions with the sequential Gauss-Newton solver [5]. This algorithm iteratively updates its estimation of the object $O(\boldsymbol{u})$ in order to match the observed images. Mathematically, we solve the following optimization problem [1, 5]:

$$\min_{O(\boldsymbol{u})} \sum_l \sum_{\boldsymbol{x}} |\sqrt{I_l(\boldsymbol{x})} - |\mathcal{F}^{-1}\{\hat{P}(\boldsymbol{u})O(\boldsymbol{u}-\boldsymbol{u}_l)\}||^2. \tag{3}$$

### 2.2. Embedded pupil recovery

An extension to FPM called embedded pupil recovery (EPRY), described in [7] and further developed in [5], involves iteratively solving for the pupil function, along with the object, as part of a joint reconstruction process. The modified optimization problem becomes:

$$\min_{O(\boldsymbol{u}),P(\boldsymbol{u})} \sum_l \sum_{\boldsymbol{x}} |\sqrt{I_l(\boldsymbol{x})} - |\mathcal{F}^{-1}\{\hat{P}(\boldsymbol{u})O(\boldsymbol{u}-\boldsymbol{u}_l)\}||^2. \tag{4}$$

The EPRY solver alternates between object and pupil updates during the iterative reconstruction. This software-only modification of FPM conveniently enables aberration estimation and digital removal without any hardware modifications. However, a drawback of joint reconstruction is that optimizing both the object and the pupil at once leads to interdependence between object and pupil reconstructions and therefore can lead to pupil-dependent object errors and vice-versa.

### 2.3. Diffuser-based characterization scheme

An alternate solution to measuring (and digitally removing) aberrations entails collecting images of a diffuser taken with different off-axis illumination, and then analyzing their power spectra [9, 17]. This method makes use of self-interference patterns formed in the power spectra of speckle images [21, 22], whose expected value depends on the aberrations and illumination conditions [9]. In many experimental settings, we are able to opportunistically find suitable diffusers, such as a holographic diffuser with index-matching oil in the visible range [9] or the surface roughness of an EUV photomask in the specific case of SHARP [17], which we simulate here. We first acquire a single defocused on-axis calibration image to recover a statistical description of the speckle via an initial low-dimensional nonlinear least-squares problem [17]. After the speckle has been characterized, we capture nine additional measurements, each with a different illumination angle, and then use a computational inverse problem to recover the Zernike aberration coefficients. This is performed by solving a nonlinear least-squares problem to minimize the error between observed measurements and those predicted by the forward model [9, 17]. The process is mathematically described by the following objective function:

$$\boldsymbol{c}^* = \underset{\boldsymbol{c}}{\operatorname{argmin}} \sum_{j=1}^{K} ||M_j - f_j(\boldsymbol{c})||_2^2 \tag{5}$$

where $M_j$ is the $j^{\text{th}}$ off-axis speckle measurement and $f_j$ is the forward model corresponding to the $j^{\text{th}}$ angle, mapping a set of input aberration coefficients $\boldsymbol{c}$ to the expected pupil interference pattern $f_j(\boldsymbol{c})$. Further details of the mathematical derivation of this method are discussed in the Appendix as well as in [9, 17].

Although this aberration measurement strategy involves capturing an extra dataset and inserting new hardware (a diffuser), it is relatively easy to implement in any system that can accommodate FPM, such as SHARP. The use of a separate set of calibration measurements for pupil recovery removes the dependency between object and pupil reconstructions. As our results demonstrate, this can provide a substantial improvement in image quality for quantitative phase retrieval in an aberrated imaging system.

## 3. Experimental results

We construct a model of the SHARP EUV microscope in MATLAB to simulate the performance of our system with ground truth datasets. Our parameters are $\lambda$ = 13.5 nm, NA = 0.33/4, and effective pixel size = 15 nm. Our analysis uses a 256 × 256 pixel patch of the imaging sensor. We model aberrations using Zernike polynomials, excluding piston and tilt. Piston corresponds to a constant phase shift, which is not recoverable from intensity-only measurements, and tilt corresponds to an image plane spatial translation, which can be described without aberrations [9]. We simulate Zernike polynomials of degree 3 with aberration magnitude in terms of root mean square (rms) wavefront error. For the purposes of these experiments, we generate aberrations with random Zernike coefficients at a specified magnitude. At each aberration magnitude, we generate 15 random aberration polynomials and test FPM reconstruction on three different samples with unique magnitude and phase content, given by resized and centered crops from sample images. For diffuser aberration characterization, we initialize with 100 random polynomials, and report the result with the lowest MSE after gradient descent optimization, consistent with previous literature [9, 17].

To compare the performance of software-only EPRY and our diffuser-based characterization scheme in order to decide whether the extra costs of the diffuser-based system are worth it, we initialize an optical system with a preset aberration magnitude and run both methods on the system. We simulate both the noise-free case and a level of Poisson shot noise corresponding to a representative exposure on SHARP ($5 \times 10^4$ photons/pixel).

In Figures 2 and 3, we present a comparison of the recovered intensity and phase for each method at varying aberration magnitude. We see that benchmarked against the uncorrected FPM recovery results, EPRY performs well when the aberrations are relatively weak, but breaks down when aberration magnitude is greater than $\lambda/10$ waves-rms. On the other hand, our diffuser-based aberration recovery consistently yields field reconstructions within 10% error of the original for a much larger set of aberration magnitudes.

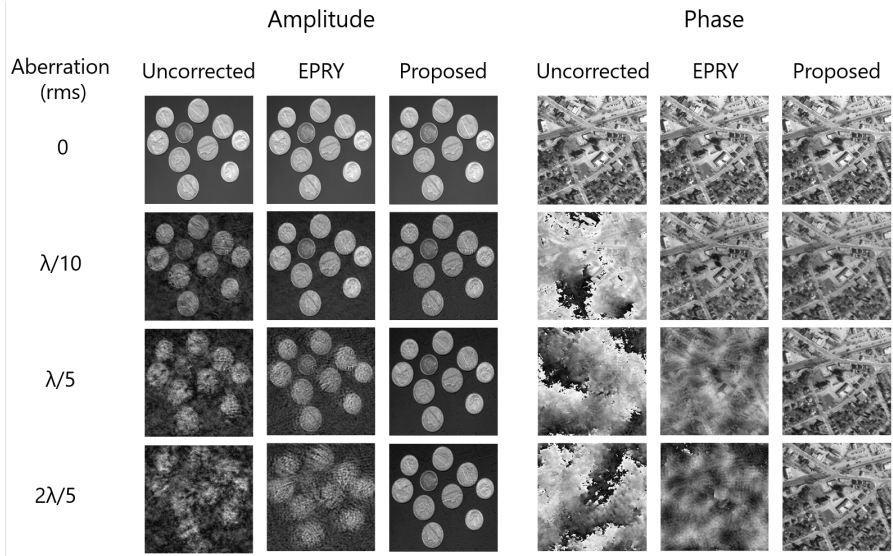

Fig. 2. Comparison of recovered amplitude and phase using no aberration correction, EPRY and the proposed method at four different aberration magnitudes (in waves-rms). At $\lambda/10$ waves-rms, the uncorrected reconstruction has significant amplitude artifacts and low-frequency phase errors (indicated by phase wrapping artifacts) while both EPRY and the diffuser-based method perform well. For aberration magnitudes greater than $\lambda/10$, EPRY reconstructions suffer from high-frequency amplitude artifacts and low-frequency phase errors, while the proposed diffuser-calibrated reconstructions remain similar to the original sample. Significant phase wrapping artifacts that exist even when EPRY is performed demonstrates the need for precalibration when performing FPM in moderately aberrated systems.

In Table 3 we also see a quantitative comparison of the image recovery error for EPRY and diffuser-based recovery. With no noise, diffuser-based recovery is able to consistently recover the image to within 10% error for aberrations up to $\lambda/3$ waves-rms, while EPRY starts failing by $\lambda/10$ waves-rms. In contrast, the uncorrected image error stays high in all cases considered. With a representative experimental photon count of $5 \times 10^4$ photon/pixel, the performance difference between these two algorithms is preserved, implying both algorithms are robust to representative levels of noise. A more detailed treatment of noise effects can be found in the Appendix.

| Aberration (waves-rms) | Noise free | | | $5 \times 10^4$ photons/pixel | |
| --- | --- | --- | --- | --- | --- |
| | Uncorrected | EPRY | Diffuser | EPRY | Diffuser |
| $\lambda/10$ | 0.97 | **0.07** | **0.05** | **0.09** | **0.07** |
| $\lambda/5$  | 0.97 | 0.35 | **0.03** | 0.38 | **0.07** |
| $\lambda/4$  | 0.99 | 0.51 | **0.03** | 0.53 | **0.08** |
| $\lambda/3$  | 0.92 | 0.59 | **0.02** | 0.60 | 0.11 |
| $\lambda/2$  | 0.93 | 0.67 | 0.16 | 0.67 | 0.36 |
| $\lambda$    | 0.96 | 0.72 | 0.99 | 0.74 | 0.96 |

Table 1. Comparison of object relative error. Under noise free imaging conditions, EPRY permits a relative reconstruction error of under 10% (bold) when the magnitude of system aberrations is under $\lambda/10$ waves-rms. This can be further extended to aberrations of magnitude up to $\lambda/3$ waves-rms with diffuser precalibration. When $5 \times 10^4$ photons/pixel of noise is modeled, we see that that performance of both algorithms is degraded, but in all cases below $\lambda/2$ the diffuser method achieves greater accuracy than EPRY.

## 4. Discussion

Compared to algorithmic self-calibration, diffuser precalibration significantly increases the maximum correctable aberration level when performing FPM. Maximum correctable aberration level is important in systems like SHARP that suffer from wavelength-scale wavefront errors. The main disadvantage of diffuser calibration is the requirement for a complete set of diffuser speckle images, both due to the extra steps needed for the data acquisition and due to the need for a speckle source. Fortunately, the latter is mitigated on SHARP as the blank part of a target photomask can be used, streamlining the process of taking diffuser calibration images.

Although aberration magnitude of the on-center 5 um patch on SHARP is known to be smaller than $\lambda/21$ waves-rms, the increased accuracy provided by diffuser precalibration at aberration magnitude between $\lambda/10$ and $\lambda/3$ waves-rms can be used to extend usable field of view. SHARP demonstrates these field-dependent aberrations, which are primarily caused by Petzval curvature, making it a system that could benefit our method's possible FOV extension. Practically, diffuser precalibration improves data throughput by making otherwise discarded off-axis patch data usable, enabling more time-efficient utilization. More generally, our method enables use of lower quality optics without loss of accuracy.

## 5. Conclusion

In this paper, we compare two different methods of aberration correction for quantitative phase imaging. Despite embedded pupil recovery (EPRY) being a simple software-only extension of the original FPM algorithm, diffuser-based aberration recovery yields higher accuracy at aberration magnitudes between $\lambda/10$ and $\lambda/3$. With aberrations of magnitude below $\lambda/10$ rms, both algorithms are able to recover the object within 10% normalized mean square error. However, for stronger aberrations, taking a separate set of calibration measurements yields more accurate aberration recovery than EPRY. We demonstrate that diffuser-based aberration recovery is a more robust method, even when shot noise is introduced into the system. Additionally, imaging systems with a large FOV often suffer from field-varying aberrations, which can severely degrade FPM reconstructions unless properly compensated. By extending correctable aberration magnitudes, diffuser calibration can extend the usable FOV since field-varying aberrations tend to become stronger with increased off-axis distance. Given that EUV photomask blanks generally satisfy the weak-phase roughness required for diffuser-based aberration recovery, we recommend this method as a competitive and straightforward method for EUV phase imaging systems with aberrations. Furthermore the technique is readily applicable to any FPM system, provided that a suitable diffuser can be found, suggesting that in many settings this could be a method to extend beyond algorithmic limits the acceptable range of aberrations under which quantitatively accurate phase imaging can be conducted.

## A. Diffuser-based aberration metrology

The model we will use is as follows:

$$\hat{I}_0(\boldsymbol{u}) = i\hat{\phi}(\boldsymbol{u}) \circ [\hat{P}^*(\boldsymbol{u}_j)\hat{P}(\boldsymbol{u}+\boldsymbol{u}_j) - \hat{P}(\boldsymbol{u}_j)\hat{P}^*(-\boldsymbol{u}+\boldsymbol{u}_j)] \quad (6)$$

where $\hat{\phi}(\boldsymbol{u})$ describes the diffuser and $\hat{I}_0$ describes the DC-suppressed counterpart of $\hat{I}$. The expression for intensity can be rearranged into a form describing the real valued forward model,

$$|\hat{I}_0(\boldsymbol{u})| = 2|\hat{\phi}(\boldsymbol{u})||\sin(S\{W(\boldsymbol{u}+\boldsymbol{u_0})\}| \quad (7)$$

where $S$ describes the symmetric decomposition operator. Our expression for the diffuser can be further broken down into two components, $\eta(\boldsymbol{u})$ is a Rayleigh distributed noise parameter and $\hat{\phi}_d(\boldsymbol{u})$ is a Gaussian function, both of which are dependent upon the properties of the photomask. This yields the following characterization of the photomask:

$$|\hat{\phi}(\boldsymbol{u})| = |\hat{\phi}_d(\boldsymbol{u})|\eta(\boldsymbol{u}) \quad (8)$$

We then collect a set of off-axis whitened measurements, which will be defined as:

$$\boldsymbol{m}_j = \frac{|\hat{I}_{0,j}|}{2 \cdot |\hat{\phi}_d|} \quad (9)$$

Finally, this stack of images can then be formulated into a nonlinear least squares inverse problem written as:

$$\boldsymbol{c}^* = \underset{\boldsymbol{c}}{\operatorname{argmin}} \sum_{j=1}^{K} ||\mathbb{1}[\mathcal{U}_j] \circ (\boldsymbol{m}_j - \mathbb{E}[\eta]|\sin(A_j\boldsymbol{c})|)||^2 \quad (10)$$

where $A_j$ denotes the map from aberration coefficients to sampled self-interference pattern, $\mathcal{U}_j$ denotes the support of the interference pattern, $\mathbb{1}$ is the indicator function, $\mathbb{E}$ is the expectation, and $K$ is the number of images. The details of the problem formulation are more thoroughly derived in previous work [17].

## B. EPRY optimizer convergence

We test optimizer convergence under conditions approximating the 6600 average photon of SHARP. Fig 3 shows that in this case Gauss-Newton optimization outperforms standard alternating projection based on a comparison of pupil recovery accuracy. We use normalized mean squared error as our accuracy metric, which is defined as:

$$E^2(m) = \frac{\sum_{\boldsymbol{u}} |S(\boldsymbol{u}) - \alpha S_m(\boldsymbol{u})|^2}{\sum_{\boldsymbol{u}} |S(\boldsymbol{u})|^2} \quad (11)$$

$$\alpha = \frac{\sum_{\boldsymbol{u}} S(\boldsymbol{u})S_m^*(\boldsymbol{u})}{\sum_{\boldsymbol{u}} |S_n(\boldsymbol{u})|^2} \quad (12)$$

At no aberration, Gauss-Newton (GN) is able to recover the pupil NMSE to within 3% error while Alternating Projection (AP) only recovers within 5% error. Aberrations at $\lambda/10$ are corrected to within 10% error by GN but only within 13% for AP. The superior performance of Gauss-Newton holds as we increase aberration levels up to $\lambda/3$, but the effects become less significant as aberration magnitude increases. Aberrations at $2\lambda/7$ are corrected to within 28% by GN compared to within 29.2% by AP and by $\lambda/3$, the difference between the algorithms has disappeared.

We further see in Fig 4 that starting for all iteration steps, Gauss-Newton optimization is more accurate than alternating projection by at least a 2% margin at each iterative step in the algorithm, reinforcing the choice to choose the Gauss-Newton optimizer.

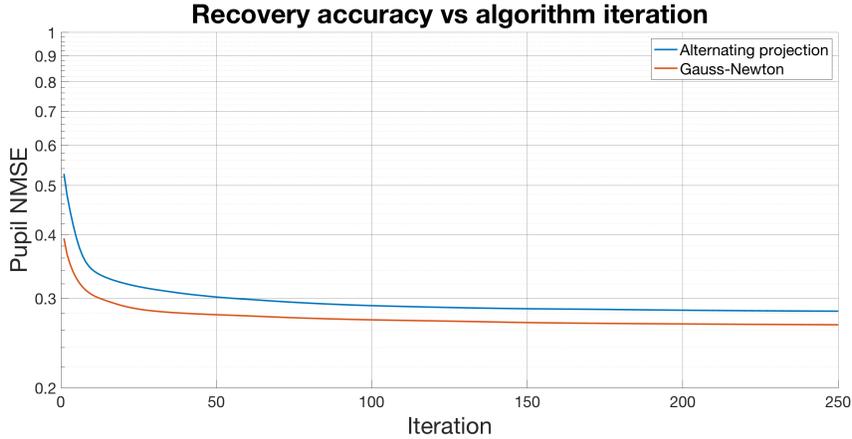

Fig. 3. Cost per iteration showing convergence of optimization methods. The example aberration shown in this plot was randomly sampled from the space of $\lambda/5$ strength. Gauss-Newton in this example both converges faster and converges to a better result.

## C. Incorporation of model noise

Realistic optical systems often suffer from Poisson shot noise, which particularly affects systems with low photon counts. Previous work shows that the choice of gradient descent method used in the FPM reconstruction highly influences the the quality of the recovered image and pupil [5]. Gauss-Newton descent was found to yield significantly better reconstruction quality than standard alternating projection when faced with Poisson noise, which is a result reinforced by our findings [2]. Second-order methods also performed well, but we feel the slight gain in accuracy is not enough to justify the increased computational cost. Since the Poisson noise found in LBNL's SHARP is on the order of 50,000 photons per pixel, photon noise is a significant effect that we decided to include in the Appendix.

Fig 5 presents a set of simulations at different aberration magnitudes, with varying degrees of Poisson noise being introduced. The number of test images and polynomials is kept the same at each aberration magnitude and noise level as in Table 1. As we decrease photon count to 50,000 photons/pixel (Fig 5b) from $10^6$ photons/pixel (Fig 5a), which is the level we see in SHARP, recovery at low aberration magnitude (below $0.3\lambda$ waves-rms) suffers. However, in the range of $0.3\lambda$ to $0.5\lambda$ waves-rms, recovery accuracy is still below 10% NMSE. EPRY performance is similar to that observed at $10^6$ photons/pixel. If we further decrease photon count to $10^3$ photons/pixel (Fig 5c), neither EPRY nor diffuser-based recovery yield significant improvements compared to the uncorrected image, but EPRY is able to converge to a better solution than diffuser-based recovery is. Importantly, even if diffuser-based recovery does not perform as well at extremely low photon counts, it is still no worse than the accuracy obtained with an uncorrected FPM system. Another takeaway is that diffuser-based recovery cannot sense very weak aberrations when there strong noise, while EPRY is not as sensitive to noise. Thus, if the target samples or detector choice is known to lead to extremely low photon counts and low aberration magnitude, diffuser-based recovery is not recommended. But as long as exposure time can be adjusted to compensate for low photon count, we recommend diffuser-based aberration recovery as a more robust and effective pre-calibration algorithm compared to existing self-calibration embedded recovery.

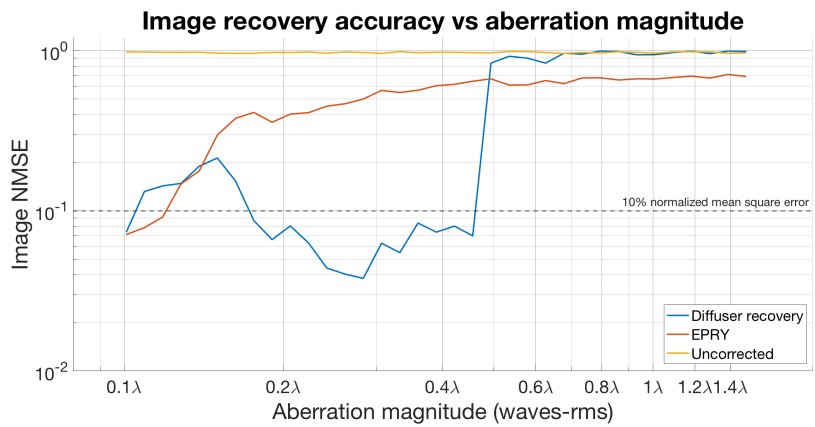

(a) $10^6$ photons/pixel

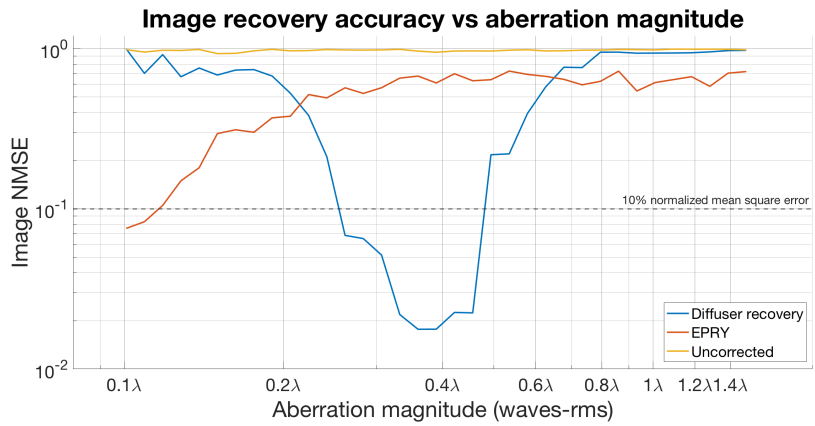

(b) 6600 photons/pixel (SHARP)

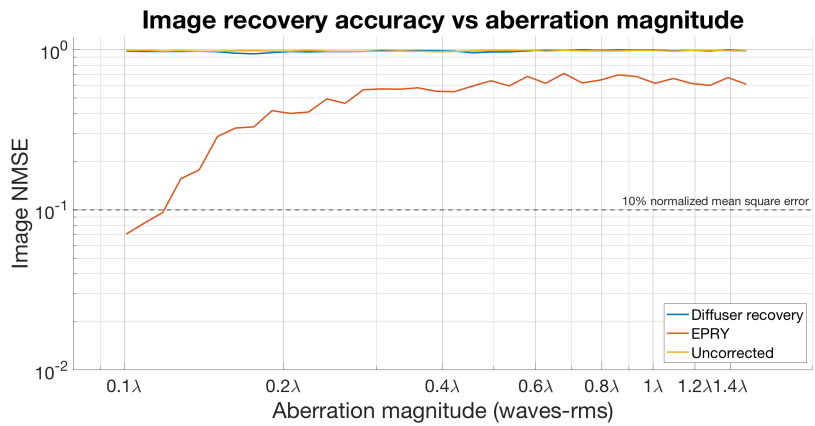

(c) $10^3$ photons/pixel

Fig. 4. Comparisons of method reconstruction error at different aberration magnitudes and photon counts.